\documentclass[prl,twocolumn,tightenlines,showpacs,nofootinbib,superscriptaddress]{revtex4-1}
\usepackage{bm,dcolumn,amsmath,graphicx,amsfonts,amssymb}

\usepackage{hyperref} 
\usepackage{xcolor}
\definecolor{blue}{rgb}{0.,0.,0.5}   
\hypersetup{
	colorlinks,
	linkcolor={blue},
	citecolor={blue},
	urlcolor={blue}
}

\newcommand{\bra}[1]{\langle #1|}	
\newcommand{\ket}[1]{|#1\rangle}	
\newcommand{\braket}[1]{\langle #1\rangle}	
\renewcommand{\v}[1]{\boldsymbol{#1}}		
\newcommand{\E}[1]{\times10^{#1}}	
\newcommand{\g}{\gamma} 
\renewcommand{\i}{\imath} 
\renewcommand{\j}{\jmath} 

\renewcommand{\P}{$\mathcal{P}$}
\newcommand{\C}{$\mathcal{C}$}
\newcommand{\T}{$\mathcal{T}$}

\newcommand{\diagram}[1]{}		
\renewcommand{\diagram}[1]{#1}	

\begin{document} 

\title{Limiting $\mathcal{P}$-odd interactions of  cosmic fields with electrons, protons and neutrons}

\author{B. M. Roberts}\email[]{b.roberts@unsw.edu.au}
\author{Y. V. Stadnik}\email[]{y.stadnik@unsw.edu.au}
\author{V. A. Dzuba}   
\affiliation{School of Physics, University of New South Wales, Sydney 2052, Australia}
\author{V. V. Flambaum} 
\affiliation{School of Physics, University of New South Wales, Sydney 2052, Australia}
\affiliation{New Zealand Institute for Advanced Study, Massey University, Auckland 0745, New Zealand}
\author{N. Leefer}
\affiliation{Helmholtz Institute Mainz, Johannes Gutenberg University, 55099 Mainz, Germany}
\author{D. Budker}	
\affiliation{Helmholtz Institute Mainz, Johannes Gutenberg University, 55099 Mainz, Germany}
\affiliation{Department of Physics, University of California at Berkeley, Berkeley, CA 94720-7300, USA}
\affiliation{Nuclear Science Division, Lawrence Berkeley National Laboratory, Berkeley, CA 94720, USA}

\date{ \today }

\begin{abstract}

We propose methods for extracting limits on the strength of $\mathcal{P}$-odd interactions of pseudoscalar and pseudovector cosmic fields with electrons, protons and neutrons.
Candidates for such fields are dark matter (including axions) and dark energy, as well as several more exotic sources described by standard-model extensions. 
Calculations of parity nonconserving amplitudes and atomic electric dipole moments induced by these fields
are performed for H, Li, Na, K, Rb, Cs, Ba$^+$, Tl, Dy, Fr, and Ra$^+$.
From these calculations and existing measurements in Dy, Cs and Tl, we constrain the
interaction strengths of the parity-violating static pseudovector cosmic field to be $7\times10^{-15}$ GeV with an electron, and $3\times10^{-8}$ GeV with a proton.

\end{abstract}
\pacs{11.30.Er, 14.80.Va, 31.15.A-, 95.35.+d} 
\maketitle

\paragraph{Introduction.---}

Among the most important unanswered questions in fundamental physics are the strong {\C\P} problem, the puzzling observation that quantum chromodynamics (QCD) does not appear to violate the combined charge-parity (\C\P) symmetry, see e.g.~\cite{Weinberg1976,*Weinberg1978,Peccei1977a,*Peccei1977b,Wilczek1978,*Moody1984},
and  dark matter and dark energy, see e.g.~\cite{Bertone2005,*Spergel2007,*Agnese2013,Riess1998,*Perlmutter1999}.
One elegant solution to the strong {\C\P} problem invokes the introduction of a pseudoscalar particle known as the axion~\cite{Peccei1977a,*Peccei1977b} (see also \cite{Kim1979,Zakharov1980,*[][{ [Sov. J. Nucl. Phys. {\bf31}, 529 (1980)].}]Zhitnitsky1980,Srednicki1981}).
It has been noted that the axion may also be a promising cold dark matter (CDM) candidate. 
Thus axions, if detected, could resolve both the CDM and strong {\C\P} problems~\cite{Kim2010}.

We consider the observable effects that arise from the {\P}-odd couplings of hypothetical pseudoscalar and pseudovector cosmic fields with atomic electrons and nuclei. 
The existence of a cosmic field that interacts with electrons or nucleons in a parity-violating manner would induce a mixing of opposite-parity states and lead to observable effects,
including parity nonconservation (PNC) amplitudes  and atomic electric dipole moments (EDMs),
the measurement of which would probe the properties of the fields that gave rise to them~\cite{Bolokhov2008,Graham2011,*Graham2013,Budker2014,LeeferCPT2013,Stadnik2014}. 
PNC amplitudes are electric dipole ($E1$) transitions between states of the same nominal parity.
Conventionally, the main contribution to these come from $Z^0$-boson exchange between the nucleus and atomic electrons, see e.g.~\cite{Khriplovich1991,FlambaumReview2004}.
Studies of atomic PNC and EDMs are relatively inexpensive low-energy tests of the standard model
that are complementary to direct tests performed at high energy.
Measurements and calculations of the Cs $6s$-$7s$ PNC amplitude stand as the most precise atomic test of the electroweak theory to date, see e.g.~\cite{Bouchiat1982,Wieman1997,DzubaCPM1989plaPNC,
Blundell1992,*KozlovCs2001,*Vasilyev2002,*DzubaCs2002,*Porsev2009,*Porsev2010,OurCsPNC2012}.

In this work, we perform calculations of cosmic-field induced PNC amplitudes and atomic EDMs for 
several atoms and ions.  
In conjunction with experimental data, these calculations are necessary for determining or placing limits on important pseudoscalar and pseudovector cosmic-field parameters. 
We combine these calculations with the results of existing PNC experiments in Cs, Tl and Dy to extract limits on the interaction strength of a static pseudovector field with electrons and protons. The same method can be directly applied to extract a limit for neutrons when appropriate experimental data become available. Such experiments, e.g.~with Yb~\cite{Tsigutkin2009,*Tsigutkin2010}, are currently under way.

\paragraph{Theory.---}

The Lagrangian densities for the interaction between fermions and a pseudoscalar (PS) field via a derivative-type and a direct pseudoscalar coupling read 
\begin{equation}
\label{eq:L-g5}
\mathcal{L}_{\g^5}^{\rm PS} = \eta\hbar\, (\partial_\mu\phi)\, \bar\psi \gamma^\mu\gamma^5\psi,%
\end{equation}
and
\begin{equation}
\label{eq:L-ig0g5}
\mathcal{L}_{i\g^0\g^5}^{\rm PS} =- i \zeta m_fc^2\, \phi  \,  \bar\psi \gamma^5  \psi,
\end{equation}
 respectively, 
where $\psi$ is the Fermion wavefunction with $\bar\psi \equiv \psi^\dagger\gamma^0$, $\eta$ and $\zeta$ are dimensionless coupling constants (into which we have absorbed the amplitude of the field),  $\g^0$   
 and $\g^5$ are Dirac matrices, and $m_f$ is the mass of the fermion. 
\diagram{The interactions (\ref{eq:L-g5}) and (\ref{eq:L-ig0g5}) are represented by the same Feynman diagram (Fig.~\ref{fig:axion_FD}).}

In the above equations, $\phi=\phi(\v{r},t)$ is the dynamic PS field in question. 
Below we show  that interactions (\ref{eq:L-g5}) and (\ref{eq:L-ig0g5}) involving a static field do not lead to atomic parity-violating effects.
With the assumption that the motion of the observer with respect to the field is slow compared to the 
speed of light,
we can express this field as 
$\phi(\v{r},t) = \cos(\omega_\phi t)$, for a particular choice of phase~\cite{Stadnik2014}, where $\hbar\omega_\phi$ is the energy of the field excitation.
The time-derivative part of the PS $\g^5$ interaction 
(\ref{eq:L-g5}), and the PS $i\gamma^0\gamma^5$ interaction (\ref{eq:L-ig0g5}), lead to the interaction Hamiltonians: 
\begin{align}
&\hat h_{\g^5}^{\rm PS} = \eta \hbar\omega_\phi  \sin(\omega_\phi t) \g^5,
\label{eq:H-g5}  \\
&\hat h_{i\g^0\g^5}^{\rm PS} = i \zeta m_fc^2 \cos(\omega_\phi t) \g^0\g^5.
\label{eq:H-ig0g5}
\end{align}

\diagram{
\begin{figure}
\begin{center}
\includegraphics[width=3.5cm]{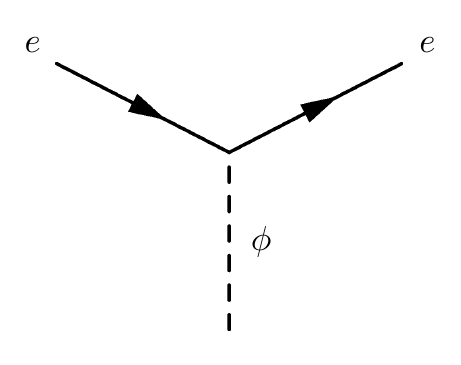}~~~~~
\includegraphics[width=3.5cm]{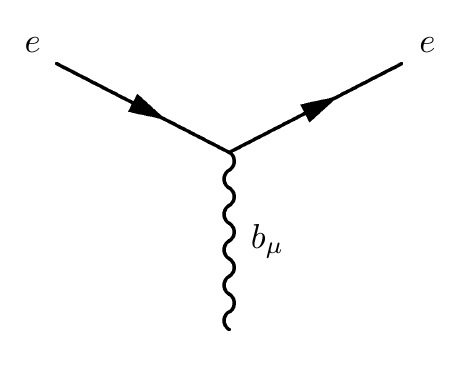}
\caption{Fundamental vertices for the interaction of an electron with a pseudoscalar cosmic field $\phi$ via the derivative-type coupling (\ref{eq:L-g5}) and the pseudoscalar coupling (\ref{eq:L-ig0g5}), and for the interaction of an electron with a pseudovector cosmic field $b_\mu$ via the  coupling (\ref{eq:L-PV}).}
\label{fig:axion_FD}%
\end{center}%
\end{figure}%
}%


The Lagrangian density for the interaction of fermions with a pseudovector (PV) cosmic field%
 \diagram{ (Fig.~\ref{fig:axion_FD})} 
is
\begin{align}
\mathcal{L}_{\rm \gamma^5}^{\rm PV} 
&= b_\mu\bar\psi \gamma^\mu\gamma^5\psi   ,
\label{eq:L-PV}
\end{align}
where 
we have absorbed the strength of the interaction into the components of the field, {${b_\mu=(b_0,-\v{b})}$}.
The time-component of this coupling leads to the interaction Hamiltonian:
\begin{equation}
\hat h_{\g^5}^{\rm PV}
= b_0(t) \g^5,
\label{eq:H-PV}
\end{equation}
which could be either static [$b_0(t)=b_0$] or dynamic [$b_0(t)=b_0\sin(\omega_b t)$].

\paragraph{Interaction with electrons.---}
The interactions (\ref{eq:H-g5}) and (\ref{eq:H-ig0g5}) with electrons induce small, oscillating contributions to  PNC amplitudes and atomic EDMs.
The effects of (\ref{eq:H-PV}) mimic those of (\ref{eq:H-g5}) in the dynamic case,  and mimic the conventional nuclear-spin-independent weak-charge ($Q_W$) induced PNC signal in the static case.

The matrix elements of the $\gamma^5$ and $ i\g^0\g^5$ operators are not entirely independent, they are related via: 
\begin{equation}
\bra{b}\sum_\i i\gamma^0_\i\gamma^5_\i\ket{a} = \frac{i}{2m_ec^2}(E_b-E_a)\bra{b}\sum_\i\gamma^5_\i\ket{a},  
\label{eq:relation}
\end{equation}
where the states $a$ and $b$ are eigenstates of the atomic Hamiltonian, $\hat H$, with eigenvalues $E_a$ and $E_b$, respectively, and the index $\i$ stands for summation over electrons.
Equation~(\ref{eq:relation}) follows directly from the identity
${i\gamma^0_\i\gamma^5_\i = \frac{i}{2m_ec^2}[\hat H,\gamma^5_\i]}$, 
which holds for the atomic Dirac-Hamiltonian including electromagnetic interactions.

We now show that a static PS interaction cannot give rise to observable {\P}-odd effects in atoms in the lowest order (though note that a static pseudo{vector} field can).
To see this for a derivative-type coupling, note that  the time-derivative term in (\ref{eq:L-g5}) vanishes for static $\phi$, and that
the spatial-derivative term in (\ref{eq:L-g5}) is proportional to $(\nabla\phi)\cdot\v{\sigma}_e$ (in the non-relativistic limit), and does not contribute to the mixing of opposite-parity atomic states.
To see this for the pseudoscalar coupling (\ref{eq:L-ig0g5}), we 
write the perturbed wavefunction for atomic state $a$ to first-order as
$
 \ket{\tilde{a}}  
=  \ket{a} - \frac{i \zeta}{2}\Sigma_\i\gamma^5_\i\ket{a},
$
where we have used the relation (\ref{eq:relation}).
Hence, the first-order correction induced by the static PS interaction (\ref{eq:L-ig0g5}) to 
the amplitude of an electromagnetic interaction,
which has the form{ $j_\mu A^\mu = \psi^\dagger_b(A^0 + \v{\alpha}\cdot\v{A})\psi_a$}, where $A^\mu=(A^0,\v{A})$ is the photon field and $\v{\alpha}$ is a Dirac matrix,
 is reduced to
$\bra{b}[ \v{\alpha},\gamma^5]\cdot\v{A}\ket{a}=0$. 
There are thus no corrections to electromagnetic amplitudes, including PNC amplitudes and atomic EDMs.

To analyze the dynamic effects, we apply first-order time-dependent perturbation theory (TDPT) --  see~Ref.~\cite{Stadnik2014} for further details. 
We now make the assumption that the energy of the field particle is much smaller than the energy separation between all opposite-parity states of interest, i.e.~$\hbar\omega_{\phi/b}\ll |E_{a,b}-E_n|$ for all $n$.
For a relatively light field particle, there is no loss of generality in making this assumption, except in the case where the atomic system of interest possesses close levels of opposite parity. This case will be investigated for Dy.

With this, we can present four comparatively simple formulae for the dynamic PNC amplitudes and atomic EDMs induced by the PS interactions 
(\ref{eq:H-g5}) and (\ref{eq:H-ig0g5}):
\begin{align}
&E_{\rm PNC}^{\rm PS}(\g^5) = {\eta \hbar\omega_\phi}\sin(\omega_\phi t) K_{\rm PNC}, \label{eq:pncg5}\\
&E_{\rm PNC}^{\rm PS}(i\g^0\g^5) = \frac{\zeta\hbar\omega_\phi}{2}\sin(\omega_\phi t) K_{\rm PNC},	\label{eq:pncg0g5}\\
&d_{\rm EDM}^{\rm PS}(\g^5) = {-2i\eta  \hbar^2\omega_\phi^2 \cos(\omega_\phi t)} K_{\rm EDM}, \label{eq:edmg5}\\
&d_{\rm EDM}^{\rm PS}(i\g^0\g^5) =  {-i\zeta \hbar^2\omega_\phi^2} \cos(\omega_\phi t) K_{\rm EDM}. \label{eq:edmg0g5}
\end{align}
For the PV interaction (\ref{eq:H-PV}), the induced PNC amplitude,
\begin{equation}
E_{\rm PNC}^{\rm PV} = b_0(t) K_{\rm PNC},
\label{eq:PVpncg5}
\end{equation}
can be either static, with $b_0(t)=b_0$, or dynamic, with $b_0(t)=b_0\sin(\omega_bt)$.
In the dynamic case, the PV interaction (\ref{eq:H-PV}) also gives rise to an oscillating atomic EDM:
\begin{equation}
d_{\rm EDM}^{\rm PV} = {-2i b_0   \hbar\omega_b \cos(\omega_b t)} K_{\rm EDM}.
\label{eq:PVedmg5}
\end{equation}
In the above equations, we have defined the henceforth named {\em atomic structure coefficients}:
\begin{align}%
&K_{\rm PNC} = \sum_{n,\i,\j} \left[ \frac{\bra{b}\v{d}_\i\ket{n}\bra{n} \g^5_\j \ket{a}}{E_a-E_n} +\frac{\bra{b}\g^5_\j\ket{n}\bra{n} \v{d}_\i \ket{a}}{E_b-E_n} \right],
\label{eq:kpnc}
\\
&K_{\rm EDM} = \sum_{n,\i,\j}  \frac{\bra{a}\v{d}_\i\ket{n}\bra{n} \g^5_\j \ket{a}}{(E_a-E_n)^2}
\label{eq:kedm},
\end{align}
where 
$\v{d}_\i=-e\v{r}_\i$ 
is the $E1$ operator, and the indices $\i$ and $\j$ stand for the summation over atomic electrons.

Formulae (\ref{eq:pncg5}--\ref{eq:PVedmg5})  provide the link between the atomic-structure calculations and the fundamental physics necessary to extract quantitative information about the fields in question.
In deriving these equations we made use of Eq.~(\ref{eq:relation}).
Notice that the atomic-structure coefficients are the same for the $\g^5$ and $i\g^0\g^5$ cases.

The $K_{\rm PNC}$ coefficients vanish in the non-relativistic limit~\cite{Stadnik2014}.
In the calculations, this leads to significant cancellation between terms in the sum~(\ref{eq:kpnc}). If the calculations were exact, this would eliminate the non-relativistic part of the amplitude and leave only the relativistic corrections, constituting the correct result.
In practice, however, the cancellation leads to significant instabilities in the calculations. 
To bypass this problem, we express the $\g^5$ operator via the exact relation:
\begin{equation}%
\g^5_\i = \frac{i}{c} [\hat H,\v{\hat\Sigma}_\i\cdot\v{r}_\i]+2\g_\i^5\hat K_\i,%
\label{eq:g5relation}%
\end{equation}%
where $\v{\hat\Sigma}$ is the Dirac spin matrix, $\hat K = -1-\v{\sigma}\cdot\v{L}$ 
[$\hat K\Omega_\kappa=\kappa\Omega_\kappa$  for the spherical spinor $\Omega_\kappa$ with $\kappa=(l-j)(2j+1)$], 
where $\hat H$ is
the atomic Dirac-Coulomb Hamiltonian, 
$\v{L}$ and $l$ are the operator and value of the orbital angular momentum, and $j$ is the total  angular momentum of the  single-electron atomic states.
The commutator in (\ref{eq:g5relation}) cancels exactly in the amplitude, and does not contribute.
We thus calculate the $K_{\rm PNC}$  coefficients free of large cancellation by using only the last term in (\ref{eq:g5relation}).


\paragraph{Interaction with nucleons.---}

The main contribution to the nuclear-spin-dependent PNC amplitude comes from the interaction of electrons with the anapole moment (AM), a {\P}-odd, {\T}-even nuclear moment that arises due to parity-violating nuclear forces~\cite{Flambaum1984}.
The interaction between a PS or PV cosmic field and an unpaired proton or neutron will give rise to a contribution to the AM.
This was considered for a PS field in Ref.~\cite{Stadnik2014}.

In the case of the static PV interaction (\ref{eq:H-PV}), the dimensionless constant quantifying the magnitude of the anapole moment can be expressed  as
$\widetilde\varkappa_{a}=\varkappa_a+\varkappa_b$, 
where $\varkappa_a$ is due to parity-violating nuclear forces, and $\varkappa_b$ is 
due to the PV field and is
related to the field parameter $b_0^{N}$ (the superscript denotes either a proton or neutron):
\begin{equation}
\label{anapole_kappa_PV}
\varkappa_b =  \frac{2\sqrt{2} \hbar\pi  \alpha\mu\braket{r^2}}{G_F m_p c } b_0^{N}, 
\end{equation}
where $m_p$ and $\mu$ are the mass and magnetic moment (in nuclear magnetons) of the unpaired nucleon, respectively, $G_F$ is the Fermi constant,
and we take the mean-square radius 
$\braket{r^2}= \frac{3}{5} r_0^2 A^{2/3}$, with $r_0 = 1.2$ fm, and $A$ the atomic mass number~\cite{Flambaum1984,Stadnik2014}.
No new atomic calculations are required -- a limit on $b_0^{N}$ can be extracted directly from existing measurements and calculations of $\varkappa_a$.
Note that interactions   with the dynamic PS (\ref{eq:H-g5}) and PV  (\ref{eq:H-PV}) fields would induce oscillating nuclear anapole and Schiff moments, which would contribute to nuclear-spin dependent PNC amplitudes and atomic EDMs, respectively~\cite{Stadnik2014}.

\paragraph{Results and discussion.---}


Apart from Dy, we treat all the considered atoms as single-valence systems.
We then use the correlation potential method to include core-valence correlations~\cite{DzubaCPM1989plaPNC}.
Core polarization and interactions with external fields are taken into account with the time-dependent Hartree-Fock method~\cite{DzubaCPM1989plaPNC}.
We estimate the uncertainty in these quantities from the effect that including correlations has on the values.

By expressing the second term on the right hand side of (\ref{eq:g5relation}) as $2\g^5\hat K=-2\g^0\g^5(\g^0\hat K)$, and noting that single-particle states are eigenstates of $\g^0\hat K$ (with eigenvalue $\kappa$), we can use Eq.~(\ref{eq:relation}) 
to invoke  the closure relation and the amplitude for single-particle states reduces to
\begin{equation}
K_{\rm PNC} = \frac{1}{m_ec^2}(\kappa_b+\kappa_a)\bra{b}\g^5\v{d}\ket{a},
\label{eq:newkpnc}
\end{equation}
which requires no summation over intermediate states, does not contain significant cancellation, and can be calculated with relatively high accuracy.

For the $K_{\rm EDM}$ coefficients, the first term on the right hand side of (\ref{eq:g5relation}) dominates the amplitude -- it scales as $1/c$ whereas the second term scales as $1/c^{3}$. 
Inserting {${\g^5\approx i/c [\hat H,\v{\Sigma}\cdot\v{r}]}$}
into (\ref{eq:kedm}),  we see that the $K_{\rm EDM}$ coefficients scale proportionally with the static dipole polarizability, with corrections of the order $(1/c)^3$.
We use this fact as a test of our calculations, and find excellent agreement using published polarizabilty values, see e.g.~\cite{Schwerdtfeger2014}.
Results of our calculations for the atomic structure coefficients $K_{\rm PNC}$ and $K_{\rm EDM}$ are presented in Table~\ref{tab:calculations}.

\begin{table}%
\centering%
\caption{Calculated PNC and EDM atomic structure coefficients for several atomic systems (a.u.).}%
\begin{ruledtabular}%
  \begin{tabular}{lldld}%
        & \multicolumn{1}{c}{Transition} & \multicolumn{1}{c}{$K_{\rm PNC}\,(i10^{-6})$} & \multicolumn{1}{c}{State} & \multicolumn{1}{c}{$K_{\rm EDM}$} \\%
\hline%
H & $1s$-$2s$				& 0.1447	& $1s$		& 0.0164\tablenotemark[1]\\
Li & $2s$-$3s$ 				& 0.219	(3)	& $2s$ 		& 0.60(1) \\ 
Na & $3s$-$4s$ 				& 0.224(4)	& $3s$ 		& 0.61(1) \\ 
K & $4s$-$5s$ 				& 0.242(4)	& $4s$ 		& 1.09(5) \\ 
 & $4s$-$3d_{3/2}$ 			& -0.307(6)	&  			&  \\ 
Rb & $5s$-$6s$ 				& 0.247(5)	& $5s$ 		& 1.22(8) \\ 
 & $5s$-$4d_{3/2}$ 			& -0.30(1)	&  			&  \\ 
Cs & $6s$-$7s$				& 0.256(5)	& $6s$ 		& 1.6(2) \\ 
 & $6s$-$5d_{3/2}$ 			& -0.22(3)	&  			&  \\ 
Ba$^+$ & $6s$-$5d_{3/2}$ 	& -0.02(1)	&  			&  \\ 
Tl & $6p_{1/2}$-$6p_{3/2}$ 	& 0.22(5) 	&$6p_{1/2}$&  0.19(3)\tablenotemark[1]\\  
Fr & $7s$-$8s$ 				&  0.253(6) 	& $7s$ 		& 1.3(2) \\   
 & $7s$-$6d_{3/2}$ 			& -0.25(3)	&  			&  \\ 
Ra$^+$ & $7s$-$6d_{3/2}$ 	& -0.08(3) 	&  			&%
  \end{tabular}%
\end{ruledtabular}%
\tablenotetext[1]{From polarizability values of H~\cite{Tang2012} and Tl~\cite{Miller2002}.}%
   \label{tab:calculations}%
  \end{table}%

The feature of Dy that makes it a particularly interesting system for the study of atomic PNC is the presence of two nearly degenerate opposite-parity states with the same total angular momentum, $J=10$, at $E = 19797.96$ cm$^{-1}$.
We use the notation $A$ for the even-parity state and notation $B$ for the odd-parity state, following Ref.~\cite{Nguyen1997}. 
The method we use for the calculations here follows closely previous calculations of PNC effects in Dy~\cite{DzubaDy2010}.
This particular configuration interaction (CI) method is described in greater detail in Ref.~\cite{DzubaVN2008}.

For Dy, it is the quantity $\bra{B} \gamma^5 \ket{A}$ that is of direct interest, since here the transition between $A$ and $B$ is measured directly~\cite{Nguyen1997}.
Due to the near-degeneracy of the levels in Dy, the first term on the right hand side of (\ref{eq:g5relation}) does not contribute, and we perform calculations using 
$2\g^5\hat K$ instead. 
To determine the uncertainty in this matrix element, we examine the effect of removing configuration states from the CI basis.
Note that in the conventional PNC case, the relevant matrix element is highly dependent on the configurations used~\cite{DzubaDy2010}.
We find, however, that this makes little difference here, meaning the $\bra{B} \gamma^5 \ket{A}$ matrix element is quite stable. 
We calculate this to be $ 0.7(2) \E{-8}\,b_0\mathrm{~a.u.}=50(20) \,b_0$ MHz.

For the static case, the PV interaction manifests itself as a contribution to the PNC amplitude of a transition between two states of the same nominal parity.
Therefore, by combining the results of the conventional ($Q_W$ induced) PNC experiments and calculations in Cs~\cite{Wieman1997,OurCsPNC2012}, Tl~\cite{Vetter1995,Kozlov2001} and Dy~\cite{Nguyen1997,DzubaDy2010},
with the calculations of the cosmic field-induced amplitudes from the present work, it is possible to extract limits on the value of the PV cosmic field coupling constant $b_0^e$ for electrons.

Using the measured value of the AM for Cs~\cite{Wieman1997,FlambaumAnM1997} and Tl~\cite{Vetter1995,Khriplovich1995}, along with the values of $\varkappa_a$  and $\varkappa_b$ from the nuclear theory~\cite{Dmitriev1997,*Dmitriev2000,*Haxton2001c,*Haxton2002} and Eq.~(\ref{anapole_kappa_PV}), 
we have extracted limits on the constant $b_0^p$ that quantifies the interaction strength of a PV cosmic field with protons (since the anapole moment in both Cs and Tl is due to an unpaired proton). 
Note that ongoing AM measurements with ytterbium will lead to a limit on the coupling to neutrons, since here the anapole moment is due to unpaired neutrons~\cite{Tsigutkin2009,*Tsigutkin2010}.
Ongoing measurements using francium~\cite{Aubin2013} will also lead to limits on the couplings to fermions.
We present both the electron and proton PV cosmic field coupling limits in Table~\ref{tab:limit}.

  \begin{table}%
    \centering%
    \caption{Limits ($1\sigma$) on the interaction strengths of a PV cosmic field with electrons ($b_0^e$) and protons ($b_0^p$) in GeV.} 
\begin{ruledtabular}%
  \begin{tabular}{llll}
        \multicolumn{2}{c}{PNC quantity}			& \multicolumn{1}{c}{$|b_0^e|$}   & \multicolumn{1}{c}{$|b_0^p|$} \\
\hline
  Cs    & $E_{\rm PNC}$($6s$-$7s$) 					&	$2\E{-14}$		&  $3\E{-8}$\\  
  Tl    &  $E_{\rm PNC}$($6p_{1/2}$-$6p_{3/2}$)    	&	$2\E{-12}$		&  $8\E{-8}$\\
  Dy    &  $\bra{A}\hat h \ket{B}$     					&	$7\E{-15}$		&  \\
\end{tabular}
\end{ruledtabular}%
    \label{tab:limit}%
  \end{table}%

For static effects, only measurements of static PNC amplitudes from conventional PNC experiments are needed to place limits on the cosmic-field parameters.
For dynamic effects, however, a different style of experiment, in which one would measure small oscillations in the PNC amplitude or atomic EDM, is needed. 
The frequency and amplitude of these oscillations would enable one to extract values for the relevant field parameters~\cite{Budker2014,Graham2011,*Graham2013,Stadnik2014}. 
For example, a determination of the frequency would provide the mass of the particle, and the amplitude of the oscillations would lead to a determination of the constants $\eta$, $\zeta$ or $b_0$.

The high sensitivity of atomic EDM experiments makes them promising for the study of the oscillating effects considered here.
Further enhancement in the sensitivity of the EDM measurements can be obtained by tuning the experiment to a specific frequency, see, e.g.~Refs.~\cite{Graham2011,*Graham2013,Budker2014,Stadnik2014}, where oscillating-EDM experiments have been recently considered. 
For example, axions with masses of $10^{-5}$ eV/$c^2$ or $10^{-9}$ eV/$c^2$, corresponding to the ``classical'' and ``anthropic'' regions~(see, e.g.~\cite{Kim2010}), would lead to oscillations with frequencies of the order of GHz and MHz, respectively. 
For the case of axions, the coherence time may be estimated from 
$\Delta \omega_a/ \omega_a\sim(\tfrac{1}{2}m_av^2/m_ac^2)\sim(v^2/c^2)$, 
where a virial velocity of $v \sim 10^{-3} c$ would be typical in our local Galactic neighbourhood, and $\omega_a\approx m_ac^2/\hbar$~\cite{Graham2011}.

The most stringent limits on the {\P}-odd interaction of the temporal component of a static PV field
with electrons, $|b_0^e| < 7 \times 10^{-15}$ GeV, 
and protons, $|b_0^p| < 3.1 \times 10^{-8}$ GeV,
come from Dy and Cs, respectively.
These limits on the temporal components, $b_0$,
 which are derived  from {\P}-odd effects, 
are complementary to existing limits on the  spatial components, $\v{b}$, derived from {\P}-even effects due to the interaction of static cosmic fields with electrons, protons and neutrons, of
$ 1.3 \times 10^{-31}$ GeV~\cite{Heckel2008}, 
$6 \times 10^{-32} $ GeV~\cite{Brown2010}
and 
$8.4 \times 10^{-34}$ GeV~\cite{Allmendinger2014},
 respectively.
For further details and a brief history on recent developments and improvements in these limits, we refer the reader to
Refs.~\cite{Kostelecky1999,Kostelecky2011a,*Kostelecky2014}.
Note that analogous oscillating {\P}-even interactions can also be sought~\cite{Graham2013,Stadnik2014}.

The prospect that atomic systems could be used as a probe for CDM has been considered in the literature, see e.g.~\cite{Pospelov2008,Graham2011,*Graham2013,
Karshenboim2011,Pospelov2013,Derevianko2013,Stadnik2014,Blum2014}.  
In addition to inducing PNC effects and EDMs, pseudoscalar fields can also give rise to other phenomena, e.g.~the axio-electric effect~\cite{Avignone2009b,Derevianko2010,DzubaPRD2010,Derbin2013},
and spin-gravity and spin-axion-momentum couplings~\cite{Kimball2013,Venema1992,Stadnik2014,Graham2013}.
Searches for cosmic-field-induced EDMs can also be performed, e.g., in solid-state, nuclear and molecular systems. 
Static electron EDM and nuclear Schiff moment experiments in ferroelectrics are discussed in Refs.~\cite{Eckel2012,Budker2006}, for instance, and solid-state systems have already been proposed for use in the detection of axionic dark matter (see e.g.~\cite{Budker2014,Beck2013}).

We have demonstrated that atomic experiments investigating the {\P}-odd effects discussed here are a viable option for searching for evidence of pseudoscalar and pseudovector cosmic fields, and for placing constraints on their interaction strengths with electrons, protons and neutrons. 
Finally, we mention that transient EDMs may also be induced by cosmic fields in the form of topological defects~\cite{Stadnik2014a}.

\paragraph{Acknowledgements.---}
The authors would like to thank 
Michael Hohensee, 
Iosif B.~Khriplovich, 
Derek Jackson Kimball, 
V.~Alan Kosteleck\'y,
Mikhail Kozlov, 
Maxim Pospelov, 
Arkady Vainshtein, 
and 
Vladimir G.~Zelevinsky
for valuable discussions.  
This research was supported in part by the Australian Research Council, by NSF grant  PHY-1068875,
and by the Perimeter Institute for Theoretical Physics. Research at the Perimeter Institute is supported by the Government of Canada through Industry Canada and by the Province of Ontario through the Ministry of Economic Development \& Innovation.
N.~Leefer was supported by a Marie Curie International Incoming Fellowship within the 7$^{\rm th}$ European Community Framework Programme.


\bibliography{references-cosmic}

\end{document}